\documentclass[twocolumn]{aastex631}

\usepackage{pifont}
\newcommand{\cmark}{\ding{51}}%
\usepackage{threeparttable, tablefootnote}
\usepackage[version=3]{mhchem}

\begin{document}

\title{An ALMA molecular inventory of warm Herbig Ae disks: 
\\ II. Abundant complex organics and volatile sulphur in the IRS~48 disk}

\correspondingauthor{Alice S. Booth} 
\email{alice.booth@cfa.harvard.edu}

\author[0000-0003-2014-2121]{Alice S. Booth} 
\altaffiliation{Clay Postdoctoral Fellow}
\affiliation{Leiden Observatory, Leiden University, 2300 RA Leiden, the Netherlands}
\affiliation{Center for Astrophysics \textbar\, Harvard \& Smithsonian, 60 Garden St., Cambridge, MA 02138, USA}


\author[0000-0002-7935-7445]{Milou Temmink}
\affiliation{Leiden Observatory, Leiden University, 2300 RA Leiden, the Netherlands}


\author[0000-0001-7591-1907]{Ewine F. van Dishoeck}
\affiliation{Leiden Observatory, Leiden University, 2300 RA Leiden, the Netherlands}
\affiliation{Max-Planck-Institut für Extraterrestrische Physik, Giessenbachstrasse 1, 85748 Garching, Germany}

\author[0009-0006-1929-3896]{Lucy Evans}
\affiliation{School of Physics and Astronomy, University of Leeds, Leeds LS2 9JT, UK}

\author[0000-0003-1008-1142]{John D. Ilee}
\affiliation{School of Physics and Astronomy, University of Leeds, Leeds LS2 9JT, UK}

\author[0000-0003-0065-7267]{Mihkel Kama}
\affiliation{Department of Physics and Astronomy, University College London, Gower Street, London, WC1E 6BT, UK}
\affiliation{Tartu Observatory, University of Tartu, Observatooriumi 1, 61602 T\~{o}ravere, Tartumaa, Estonia}

\author[0000-0001-5849-577X]{Luke Keyte}
\affiliation{Department of Physics and Astronomy, University College London, Gower Street, London, WC1E 6BT, UK}

\author[0000-0003-1413-1776]{Charles J. Law}
\altaffiliation{NASA Hubble Fellowship Program Sagan Fellow}
\affiliation{Department of Astronomy, University of Virginia, Charlottesville, VA 22904, USA}

\author[0000-0003-3674-7512]{Margot Leemker}
\affiliation{Leiden Observatory, Leiden University, 2300 RA Leiden, the Netherlands}

\author[0000-0003-2458-9756]{Nienke van der Marel}
\affiliation{Leiden Observatory, Leiden University, 2300 RA Leiden, the Netherlands}

\author[0000-0002-7058-7682]{Hideko Nomura}
\affiliation{Division of Science, National Astronomical Observatory of Japan, 2-21-1 Osawa, Mitaka, Tokyo 181-8588, Japan}

\author[0000-0003-2493-912X]{Shota Notsu}
\affiliation{Department of Earth and Planetary Science, Graduate School of Science, The University of Tokyo, 7-3-1 Hongo, Bunkyo-ku, Tokyo 113-0033, Japan}
\affiliation{Department of Astronomy, Graduate School of Science, The University of Tokyo, 7-3-1 Hongo, Bunkyo-ku, Tokyo 113-0033, Japan}
\affiliation{Star and Planet Formation Laboratory, RIKEN Cluster for Pioneering Research, 2-1 Hirosawa, Wako, Saitama 351-0198, Japan}

\author[0000-0001-8798-1347]{Karin Öberg}
\affiliation{Center for Astrophysics \textbar\, Harvard \& Smithsonian, 60 Garden St., Cambridge, MA 02138, USA}

\author[0000-0001-6078-786X]{Catherine Walsh}
\affiliation{School of Physics and Astronomy, University of Leeds, Leeds LS2 9JT, UK}

\begin{abstract}
The Atacama Large Millimeter/submillimeter Array (ALMA) can probe the molecular content of planet-forming disks with unprecedented sensitivity. These observations allow us to build up an inventory of the volatiles available for forming planets and comets. Herbig Ae transition disks are fruitful targets due to the thermal sublimation of complex organic molecule (COM) and likely \ce{H_2O}-rich ices in these disks. The IRS~48 disk shows a particularly rich chemistry that can be directly linked to its asymmetric dust trap. Here, we present ALMA observations of the IRS~48 disk where we detect 16 different molecules and make the first robust detections of \ce{H_2^{13}CO}, \ce{^{34}SO}, \ce{^{33}SO} and \ce{c-H_2COCH_2} (ethylene oxide) in a protoplanetary disk. All of the molecular emissions, aside from CO, are co-located with the dust trap and this includes newly detected simple molecules such as \ce{HCO^+}, \ce{HCN} and \ce{CS}. Interestingly, there are spatial offsets between different molecular families, including between the COMs and sulphur-bearing species, with the latter being more azimuthally extended and radially located further from the star. The abundances of the newly detected COMs relative to \ce{CH_3OH} are higher than the expected protostellar ratios, which implies some degree of chemical processing of the inherited ices during the disk lifetime. These data highlight IRS~48 as a unique astrochemical laboratory to unravel the full volatile reservoir at the epoch of planet and comet formation and the role of the disk in (re)setting chemical complexity.

\end{abstract}


\section{Introduction} \label{sec:intro} 

Due to the sensitivity of the Atacama Large Millimeter/submillimeter Array (ALMA) we now have unmatched access to the volatile reservoir in planet-forming disks. In recent years, ALMA has enabled the detection of both new disk molecules including \ce{SO_2} and \ce{CH_3CN}, and rare isotopologues (e.g., \ce{^{13}C^{17}O} and \ce{HC^{18}O^+}) \citep{2015Natur.520..198O, 2019ApJ...882L..31B, 2021A&A...651L...6B, 2022ApJ...926..148F}. What is particularly exciting is the detection of complex organic molecules (COMs) which are defined as molecules containing at least 6 atoms and of which at least one is carbon \citep{2009ARA&A..47..427H}. Although the first detection of the simplest COM \ce{CH_3OH} in a Class II T-Tauri disk (TW~Hya) traced a very low abundance of cold \ce{CH_3OH} \citep{2016ApJ...823L..10W} subsequent observations of warmer Herbig Ae transition disks have revealed abundant thermally desorbed \ce{CH_3OH} and even other COMs of higher complexity \citep{2021NatAs...5..684B, 2021A&A...651L...5V, Booth2023, 2022A&A...659A..29B}. 
The detection of abundant COMs in warm Herbig Ae disks is clear evidence for the inheritance of ices from the earlier stages of star formation. This is because \ce{CH_3OH} only forms efficiently on the surfaces of cold dust grains and primarily via the hydrogenation of CO ice \citep{Watanabe2002, 2009A&A...505..629F, 2022ApJ...931L..33S}. 

In the warm young F/Herbig Ae disks HD~100546, IRS~48 and HD~169142 there is no evidence for significant CO freeze-out meaning that the observed reservoir of \ce{CH_3OH} cannot have formed \textit{in-situ}. This was shown directly for the HD~100546 disk using astrochemical models \citep{2021NatAs...5..684B}. Therefore, in order for \ce{CH_3OH} to be present in these systems \ce{CH_3OH} rich ices must survive the star formation process and be transported to the inner disk where they thermally sublimate. \ce{CH_3OH} will come off the grains at a similar temperature as \ce{H_2O} \citep{2022ESC.....6..597M} and therefore the bulk of the volatile content of the disks should also be in the gas phase in this region of the disk. These sources therefore give us a window into a typically unobservable molecular reservoir in disks.

The disk most rich in COMs and potentially \ce{H_2O}-derived volatiles like SO is the disk around the young star IRS~48. The IRS~48 disk has been well studied with ALMA and hosts the most asymmetric dust trap yet discovered at a distance of 60~au from the central star \citep{vanderMarel2013, 2021AJ....161...33V, 2023ApJ...948L...2Y}. Its gas mass of only 
5.5$\times10^{-4}$~M$_{\odot}$ 
is much lower than other Herbig Ae disks, yet it is very line rich, with detection's of the CO isotopologues \ce{^{12}CO}, \ce{^{13}CO}, \ce{C^{18}O}, \ce{C^{17}O} along with \ce{SO}, \ce{SO_2}, \ce{^{34}SO_2}, \ce{NO}, \ce{H_2CO}, \ce{CH_3OH}, \ce{CH_3OCH_3} and, tentatively \ce{CH_3OCHO} \citep{vanderMarel2013, 2021A&A...651L...5V, 2021A&A...651L...6B, 2022A&A...659A..29B, 2023arXiv230300768L}. 
The significance of the reported non-detections of CS, \ce{C_2H},  \ce{CN} in IRS~48 were quantified by \citet{2021A&A...651L...6B} and \citet{2023arXiv230300768L} and indicate a C/O ratio in the disk gas that is significantly less than 1. This low C/O and lack of \ce{C_2H} is consistent with \ce{H_2O} being in the gas-phase and a general lack of volatile depletion at least at the location of the dust trap \citep{2023arXiv230300768L}. 

There are several key simple molecules that have yet to be targeted in the IRS 48 disk, which would allow for a more complete comparison to other Herbig Ae disks. Here we present the results of an ALMA line survey of the IRS~48 disk where we target $>$20 molecular species. These data provide key constraints on the abundances of \ce{HCO+}, HCN, CN, \ce{C_2H} and CS in this system. Additionally, we further unravel the volatile sulphur and complex organic reservoir of the disk and discuss the physical/chemical origin of the molecular sub-structures observed.  We particularly make a direct comparison between the molecular inventory of the IRS~48 and HD~100546 disks where the initial results for the latter are presented in \citet{Booth2023_hd100546} and, contextualise the detections of COMs in these systems with protostellar environments. 

\section{Observations} \label{sec:methods}

IRS~48 was observed in the ALMA program 2021.1.00738.S (PI. A. S. Booth) and the general properties of the IRS~48 system are listed in Table~\ref{tab:A1}. 
The data consist of two spectral settings with four spectral windows each at a spectral resolution of 976.6~kHz (0.84~km~s$^{-1}$ at 350~GHz) and a bandwidth of 1.875~GHz. These spectral windows are centered at 338.790824, 340.732413, 348.916936 and 350.775389~GHz for setting A and, 344.240980, 3459.40999, 354.367095 and 356.067114~GHz for setting B. Further details on the individual execution blocks are provided in the Appendix in Table~\ref{tab:A3} and for full details the data reduction, observational set-up and imaging please refer to the companion paper which also presents data on the HD~100546 system \citet{Booth2023_hd100546}. 
The self-calibration was performed on the IRS~48 continuum data after flagging the strong lines which resulted in a continuum signal-to-noise increase from $\approx$475 to $\approx$3220. This process consisted of four rounds of phase-calibration and one round of amplitude calibration and resulted in the detection of the weak millimetre emission in the north of the IRS~48 disk. The data were imaged in CASA using \texttt{tCLEAN} with the multiscale deconvolver with a uniform velocity resolution of 0.9~km~s$^{-1}$. These $\approx$0\farcs3 data have a beam area 2.5$\times$ smaller than that presented the series of papers from \citet{2021A&A...651L...5V, 2021A&A...651L...6B, 2022A&A...659A..29B, 2023arXiv230300768L}. Individual lines were cleaned with Keplerian masks down to 4$\times$ the rms of the dirty image where the Keplerian masks were constructed using the properties for the IRS~48 disk as listed in Table~\ref{tab:A1}. The properties of the transitions imaged and the resulting beam sizes and rms noise for each line are listed in Tables~\ref{tab:lines} and \ref{tab:irs48_images} in the Appendix.

\begin{table*}
    \centering
    \caption{Properties of the IRS~48 and HD~100546 star and disk systems.}
    \begin{tabular}{c c c c c c c c c c c c c c} \hline \hline 
        Source  & Type   & Dist. & Incl. & PA  & L & M$_{*}$ & M$_{dust}$ & M$_{gas}$ & log$_{10}(\dot{M}_{acc}$) & log10($L_{Xray}$) & v$_{sys}$ & Refs.\\   &   & (pc) & (deg) & (deg) & ($L_{\odot}$) &  ($M_{\odot}$) & ($M_{\odot}$) &($M_{\odot}$) &  ($M_{\odot}$ yr$^{-1}$) & (erg~s$^{-1}$) & (km s$^{-1}$) & \\  
        \hline 
        IRS 48  & A0  & 135 & 50.0 & 100.0 & 14.3 & 2.0  &  1.5$\times10^{-5}$     & 5.5$\times10^{-4}$  & -8.40 & $<$27.0 & 4.55 & [1-7] \\
          HD 100546  &  A0-A1  & 110  & 41.7 & 146.0 & 23.5  & 2.2 & 1.1$\times10^{-3}$ & 1.5$\times10^{-1}$&  -6.81 & 28.1 &  5.70  & [8-13]\\ \hline
    \end{tabular}    
    \tablecomments{
    References:  
    [1] \citet{2012ApJ...744..116B},
    [2] \citet{Follette2015},
    [3] \citet{Gaia2018},
    [4] \citet{Bruderer2014},
    [5] \citet{vanderMarel2016-isot},
    [6] \citet{2013ApJ...769...21S},
    [7] \citet{2023arXiv230300768L},
    [8] \citet{2018A&A...620A.128V} ,
    [9] \citet{2021A&A...650A.182G}, 
    [10] \citet{2014ApJ...791L...6W}, 
    [11] \citet{2017A&A...607A.114W}, 
    [12] \citet{Kama2016twhya}, 
    [13] \citet{2012A&A...544A..78M}
}
    \label{tab:A1}
\end{table*}

\begin{figure*}
    \centering
    \includegraphics[trim={2.5cm 0cm 2.5cm 0cm}, clip,width=\hsize]{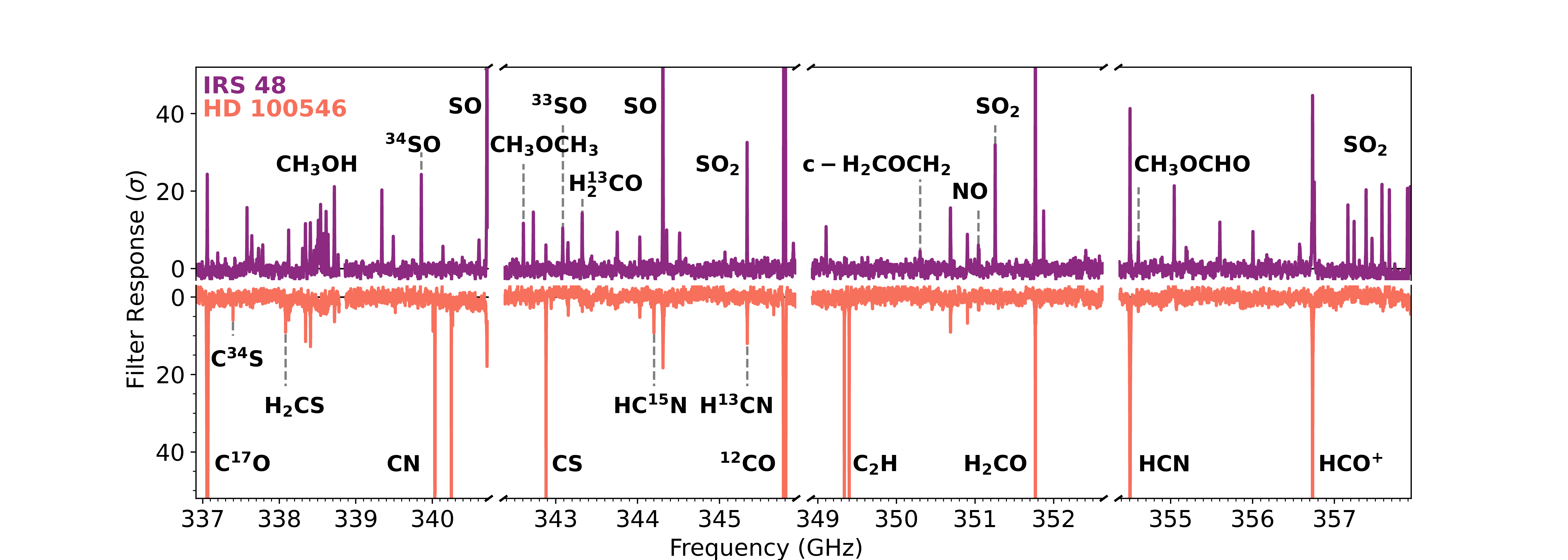}
    \caption{Matched filter responses for the IRS~48 and HD~100546 (taken from \citealt{Booth2023_hd100546}) disks showing the full frequency coverage of the observations and highlighting the main molecules detected in each disk. Note that the HD~100546 response has been inverted and the lines reaching the top and bottom of the y-axis have responses $>$50$\sigma$. The molecule labels in the top and bottom of the plot indicate from which disk the line is more strongly detected and the vertical grey lines show the location of particular molecular transitions in more crowded regions of the spectrum. Both matched filter responses were generated using Keplerian models with an outer radius of 150~au for IRS~48 and 300~au for HD~100546. A fully annotated version of the IRS~48 response is shown in Figure~\ref{fig:irs48_filter}.}
    \label{fig:1}
\end{figure*}

\section{Results}

\subsection{Molecules detected}

We use matched filtering to make an initial line identification \citep{Loomis2018}. 
This technique uses the predictable Keplerian rotation of the disk gas to detect molecular lines in the visibility data via cross correlation of the uv-data with a filter. This filter can be a smooth model, e.g. a Keplerian mask, the FITS output of a line radiative transfer model or, a strong line detection in the disk. 
In Figure~\ref{fig:1} we present the resulting matched filter response over the full data set for the IRS~48 disk with a Keplerian model with an outer radius of 150~au compared to the HD~100546 response (outer radius of 300~au) that is presented in \citet{Booth2023_hd100546}.  
HD~100546 was observed in the same manner as IRS~48 and we find that the IRS~48 disk is more line rich but there are different molecules detected in each disk.
These differences may be attributed to different physical properties of the systems and/or the dominant chemical processes. Both are disks around young A-type stars and the characteristics of these two systems are compared in Table~\ref{tab:A1}. In Section 4.2 we discuss the similarities and differences both physically and chemically between the two disks. 
The fully annotated version of the IRS~48 filter response is shown in the Appendix in Figure~\ref{fig:irs48_filter}.
From this we have detected 16 molecular species in IRS~48 disk where a detection is defined as a matched filter response of at least 4~$\sigma$. This includes robust detections of the rare isotopologues \ce{H_2^{13}CO}, \ce{^{34}SO} and \ce{^{33}SO} and, the detection of the first heterocycle -  ethylene oxide (\ce{c-H_2COCH_2}) - in protoplanetary disks. We detect two lines of \ce{c-H_2COCH_2} with the fiducial Keplerian model filter at rest frequencies of 338.7720826~GHz and 350.3036524~GHz. Using alternative image filters does not yield a significant improvement in the detection strength - likely due to the compact nature of the emission. In the channel maps compact emission from \ce{c-H_2COCH_2} is detected at the 4~$\sigma$ level over the 3 consecutive channels where the \ce{CH_3OH} lines are strongest for both lines. Interestingly, although
the isomer acetaldehyde (\ce{CH_3CHO}) us typically more abundant \citep[][]{2001ApJ...560..792I, 2017A&A...597A..53L}, it is not detected in the IRS~48 disk.
Dimethyl ether (\ce{CH_3OCH_3}) is detected again, as reported by \citet{2022A&A...659A..29B}, and their weak detection of methyl formate (\ce{CH_3OCHO}) is clearly confirmed in our data. 
A investigation into other COMs lines covered in these data and upper-limits on other non-detections will follow in Kipfer et al. (in prep). 
A summary of the molecules detected/not-detected in both the IRS~48 and HD~100546 disks are shown in Table~\ref{tab:detections}. It is unclear from visual inspection of the data if \ce{H^{13}CN} is detected in IRS~48 or not as this line is blended with a strong \ce{SO_2} line.  Using matched filtering and the HCN as a mask we find that \ce{HC^{15}N}, CN and \ce{C_2H} are all not detected. 
 

\begin{table}
\centering
\caption{Molecules detected (\cmark) and not detected (-) in the ALMA observations of the IRS 48 and HD~100546 disks presented in this paper and \citet{Booth2023_hd100546}.} 
    \begin{tabular}{c c c}
    \hline \hline 
       Molecule      &  HD 100546 & IRS 48  \\ \hline 
        \ce{^{12}CO}    & \cmark  &    \cmark  \\
        \ce{C^{17}O}    & \cmark  &    \cmark  \\
        \ce{HCO^+}      & \cmark  &    \cmark  \\
        \ce{HC^{18}O^+} & -       &   -  \\
        \ce{CN}         & \cmark  &    -  \\
        \ce{HCN}        & \cmark  &    \cmark  \\
        \ce{H^{13}CN}   & \cmark  &    ?  \\
        \ce{HC^{15}N}   & \cmark  &    -  \\
        \ce{NO}         & \cmark  &    \cmark  \\
        \ce{HC_3N}      & -       &    -  \\
        \ce{CH_3CN}     & -       &    -  \\
        \ce{C_2H}       & \cmark  &    -  \\
        \ce{c-C_3H2}    & -       &    -  \\
         \ce{CS}         & \cmark  &    \cmark  \\
        \ce{C^{34}S}    & \cmark  &    -  \\
        \ce{SO}         & \cmark  &    \cmark  \\
        \ce{^{34}SO}    & \cmark  &    \cmark  \\
        \ce{^{33}SO}    & -       &    \cmark  \\ 
        \ce{SO_2}       & \cmark  &    \cmark  \\
        \ce{OCS}        & -       &    -  \\
        \ce{H_2CS}      & \cmark  &    -  \\
        \ce{H_2CO}      & \cmark  &    \cmark  \\
        \ce{H_2^{13}CO} & \cmark  &    \cmark  \\
        \ce{CH_3OH}     & \cmark  &    \cmark  \\
        \ce{CH_3OCHO}   & \cmark  &    \cmark  \\ 
        \ce{CH_3OCH_3}  & -       &    \cmark  \\ 
        \ce{c-H_2COCH_2}& -       &    \cmark \\
        \hline
    \end{tabular}
    \tablecomments{The presence of \ce{H^{13}CN} in IRS~48 is unclear (indicated with "?") due to line blending with \ce{SO_2}.}
    \label{tab:detections}
\end{table}

\subsection{Integrated intensity maps}

\begin{figure*}
    \centering
    \includegraphics[trim={0cm 0cm 0cm 0cm}, clip,width=\hsize]{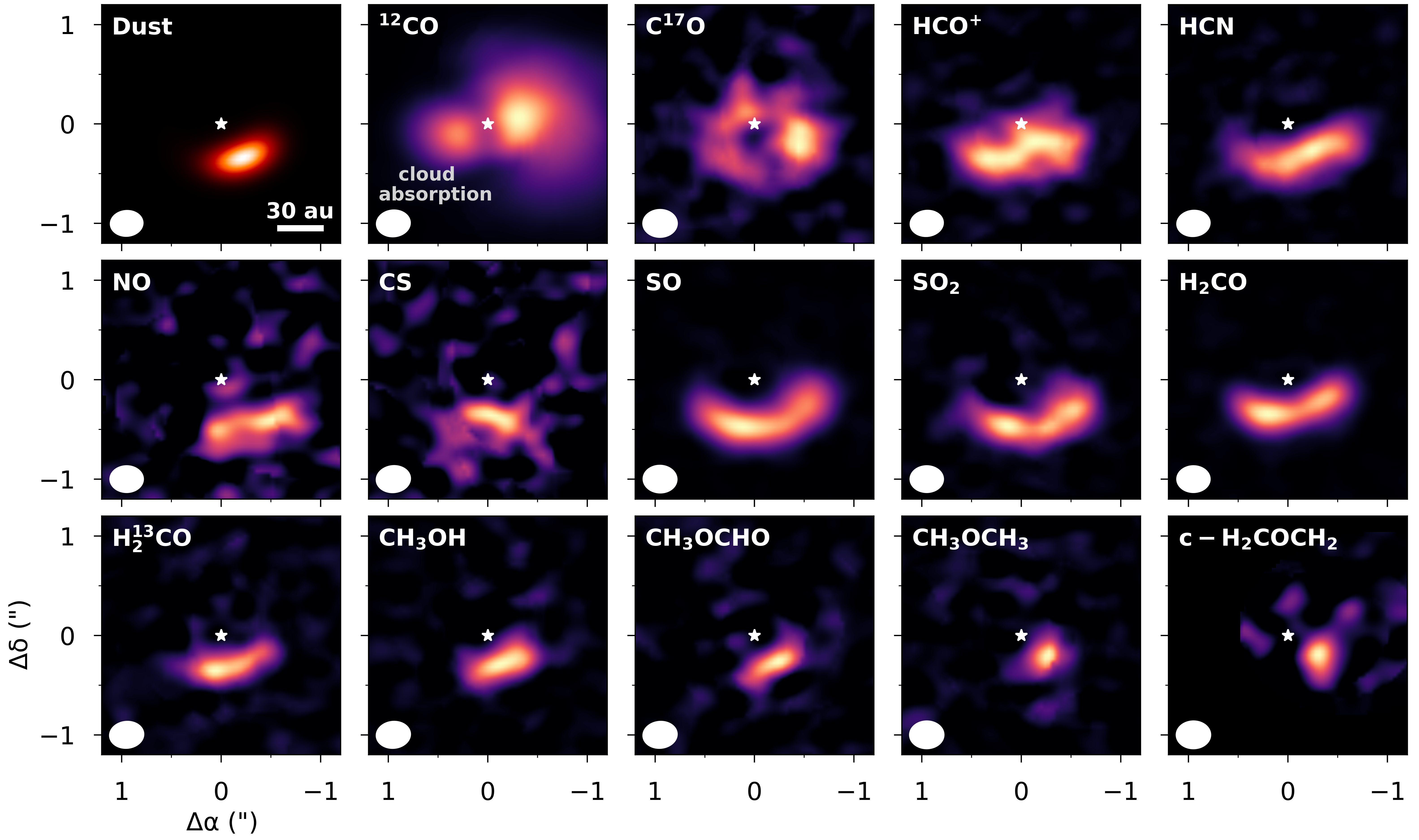}
    \caption{Integrated intensity maps of the 0.9~mm dust continuum emission and molecular line emission from the IRS~48 disk. The continuum map is shown on a color log scale to highlight the weak millimetre emission in the north of the disk. The beam is shown in the left-hand corner of each panel.}
    \label{fig:mom02}
\end{figure*}

Figure~\ref{fig:mom02} presents the 0.9~mm continuum map and the integrated intensity maps of the representative transitions of each molecule detected in the IRS~48 disk. This galley does not include the isotopologues of SO which will be the focus of a future work. These line maps were generated using the Keplerian masks generated in the CLEANing with no clipping thresholds. All of the molecules aside from \ce{^{12}CO} and \ce{C^{17}O} only show significant emission in the south of the disk - the same region of the disk as the millimetre dust trap. In the north of the disk the \ce{^{12}CO} emission suffers from cloud absorption along the minor axis of the disk but there is weak millimetre dust and \ce{C^{17}O} emission present here (also seen by \citet{2014A&A...562A..26B} in the \ce{C^{17}O} $J=6-5$). Previous studies have shown asymmetric emission for \ce{SO}, \ce{SO_2}, \ce{NO} and several of the large organics \citep{2021A&A...651L...5V, 2021A&A...651L...6B, 2022A&A...659A..29B}. Here, we present the first detections of the simple molecules \ce{HCO^+}, \ce{HCN} and \ce{CS}, and interestingly find that they all show a similar asymmetric emission morphology. However, not all of the molecules have the exact same asymmetric morphology. 

\subsection{Sub-structures in the IRS~48 disk}

\begin{figure*}
    \centering
    \includegraphics[width=0.9\hsize]{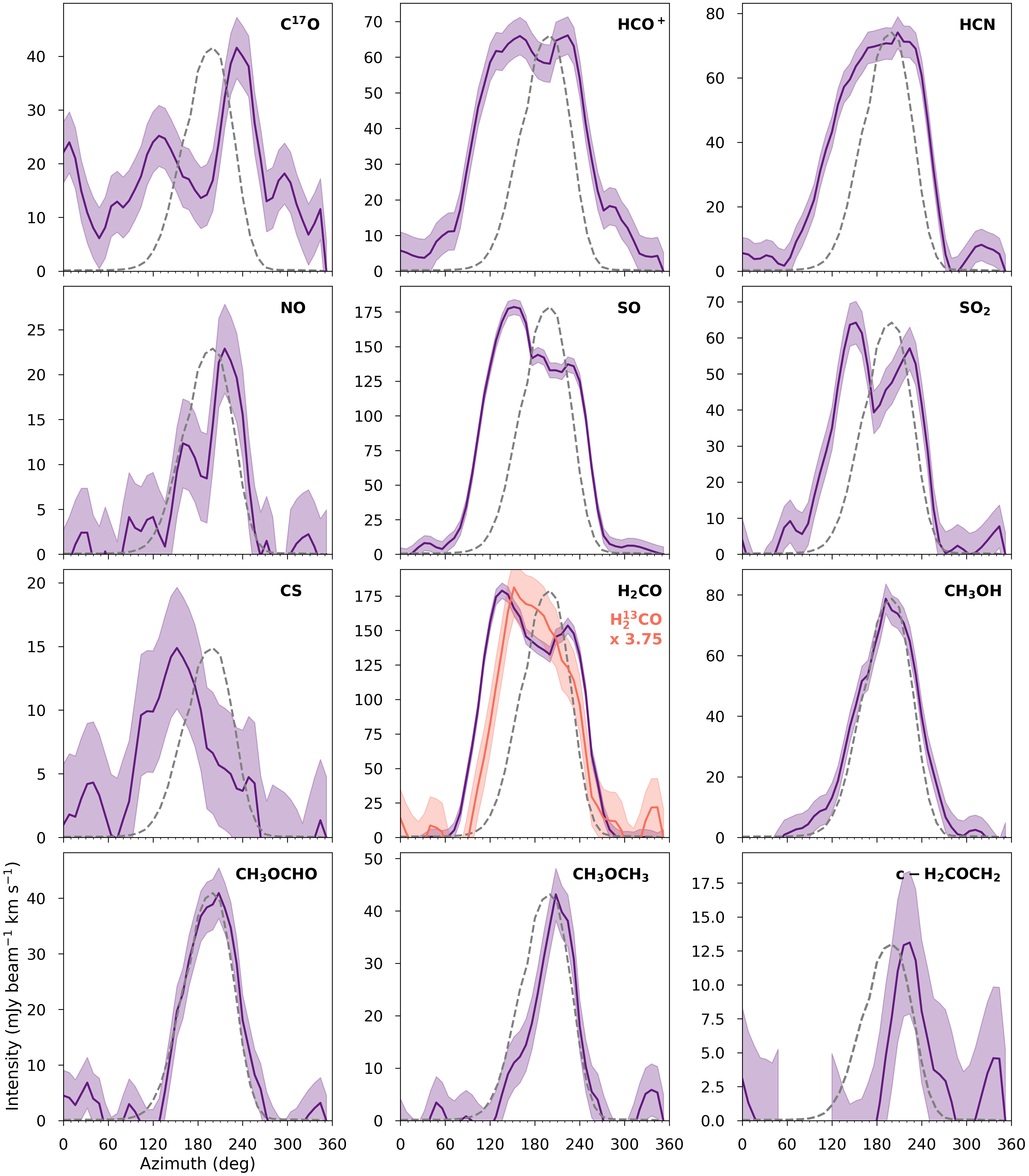}
    \caption{Azimuthal line emission profiles for the IRS~48 disk generated from the maps presented in Figure~\ref{fig:mom02}. The dashed lines show the millimeter dust emission normalised to the peak of the line emission in each panel.}
    \label{fig:irs48_azimuthal_profiles}
\end{figure*}

\begin{figure*}
    \centering
    \includegraphics[width=\hsize]{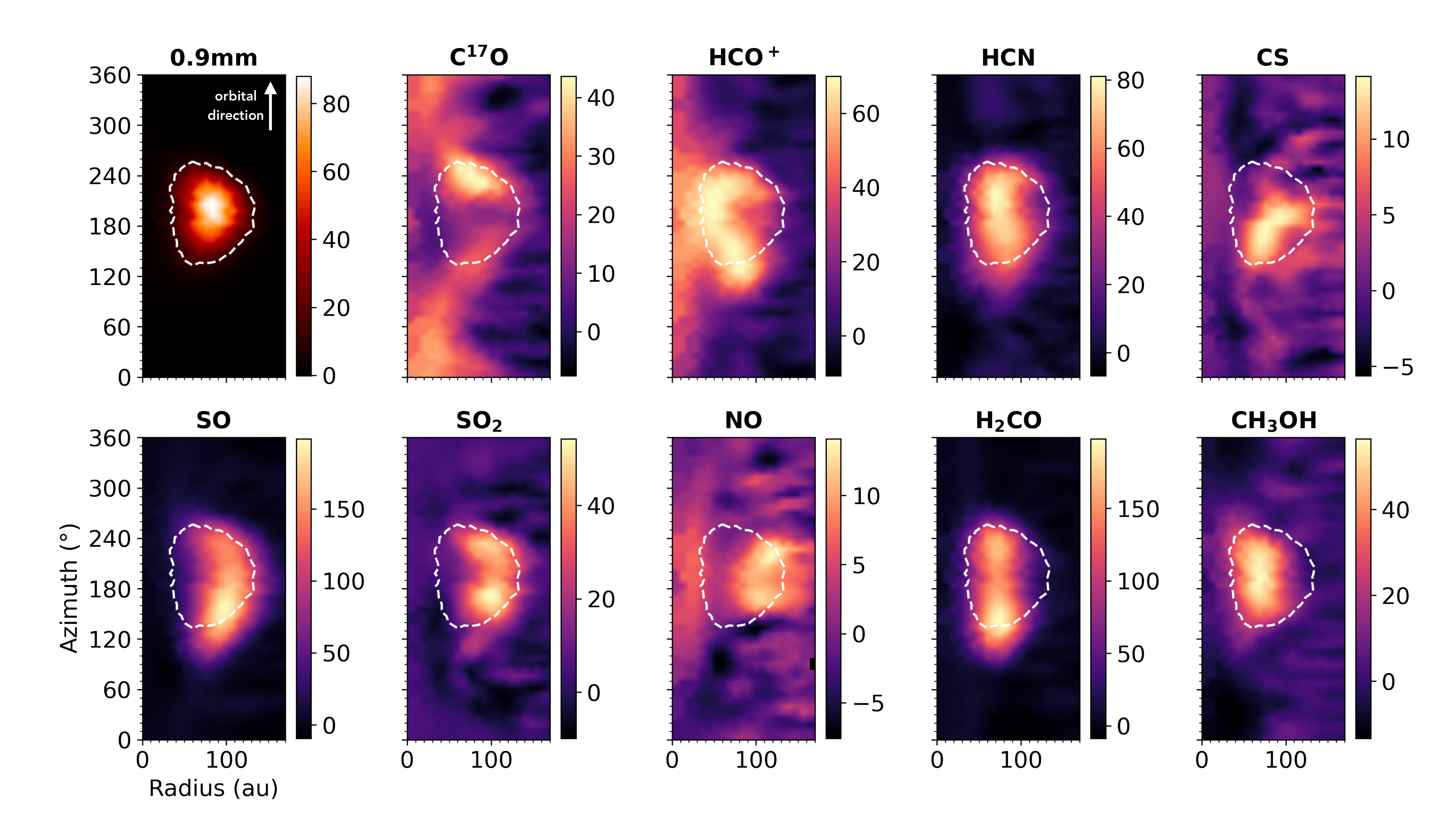}
    \caption{Polar de-projection of the IRS~48 integrated intensity maps which highlight the different emission morphologies compared to the dust. 
    The dashed contour traces the 500$\sigma$ level of the dust continuum emission. The units of the color bar are mJy~beam$^{-1}$~km~$\mathrm{s^{-1}}$ for the molecular lines and mJy~beam$^{-1}$ for the continuum. The arrow highlights the direction of the disk rotation.}
    \label{fig:irs48_azimuth}
\end{figure*}

The only molecule detected in the north of the IRS~48 disk is CO, while all of the other species are located in the south, but there are variations in where the different molecules peak both radially and azimuthally. Figure~\ref{fig:irs48_azimuthal_profiles} shows azimuthal profiles taken from the intensity maps in Figure~\ref{fig:mom02} at the radius where each of the molecules peak along with the normalised azimuthal profile of the millimetre dust. From this, it is clear that there are dips in the intensity of most species at the azimuthal peak of the dust emission. This could be due to line suppression from the optically thick dust \citep[e.g.,][]{2018ApJ...853..113W, 2020ApJ...896L...3D} but interestingly this is not as apparent for the COMs emission. The COMs emission is also significantly narrower in azimuthal extent than the simpler molecules that are detected and are located at the dust peak with a similar width to the millimetre dust. This is highlighted further in Figure~\ref{fig:irs48_azimuth} which presents a polar deprojection of the intensity maps. It is clear that the SO and \ce{SO_2} peak radially further out in the disk than the \ce{CH_3OH} and \ce{H_2CO}, which was not clear in the lower spatial resolution data presented in \citet{2021A&A...651L...5V} and \citet{2021A&A...651L...6B}. Furthermore, the \ce{HCO^+} emission is peaking closer to the star, in the gas cavity, than the \ce{CH_3OH}, and the \ce{HCN} is approximately co-spatial with the \ce{H_2CO}. The possible physical and chemical explanations for these different emission morphologies will be discussed further in Section~4.1.

\subsection{Disk integrated line fluxes}

\begin{figure*}
    \centering
    \includegraphics[trim={0cm 0cm 0cm 0cm}, clip,width=\hsize]{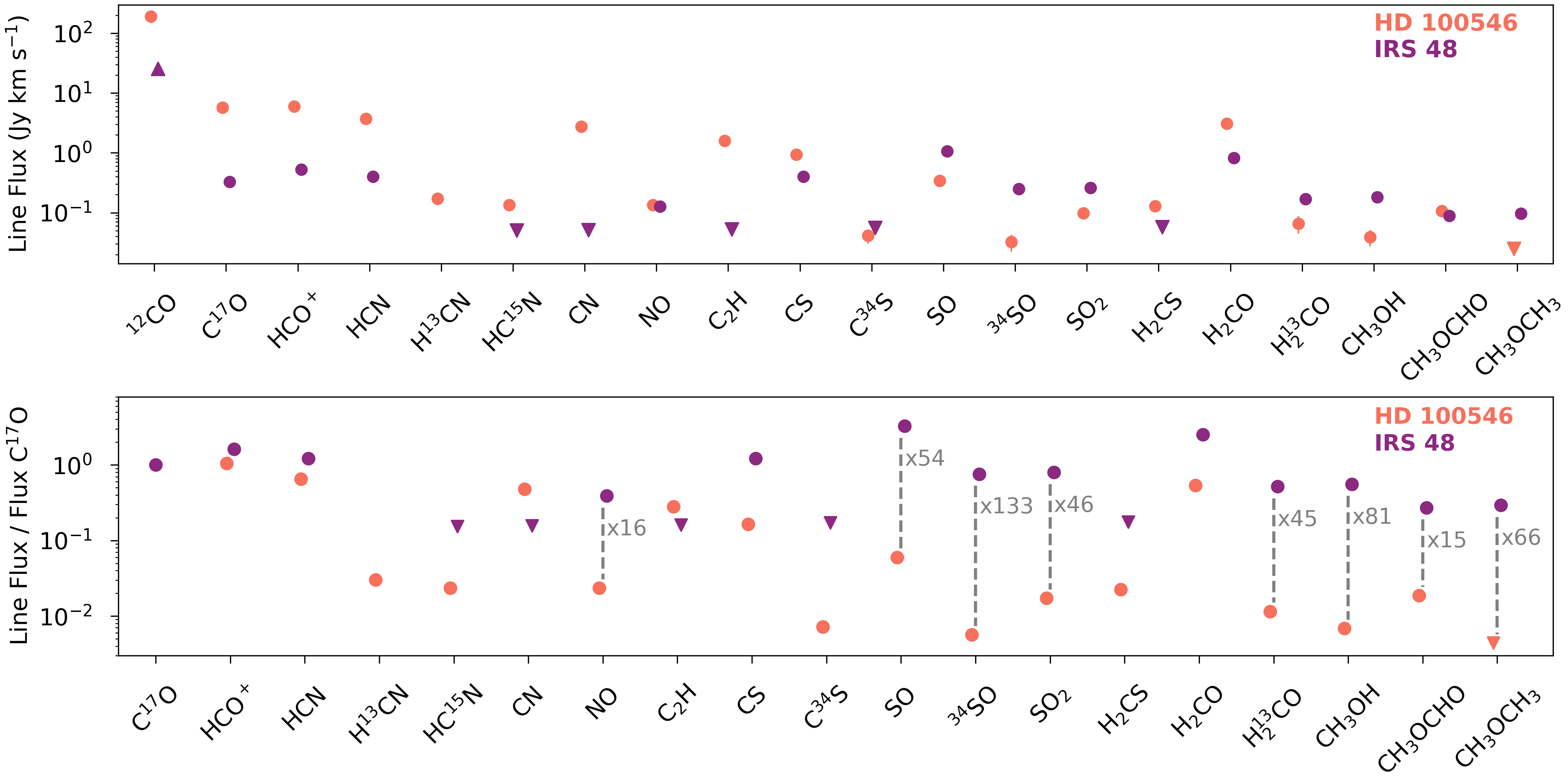}
    \caption{Top: Disk integrated fluxes for the molecules detected in the IRS~48 (purple) and HD~100546 (orange) disks. Bottom: Disk integrated fluxes relative to the \ce{C^{17}O} $J=3-2$ line flux for each disk. Vertical lines and numbers show the relative differences in the different line ratios between each disk where this value is $>$10$\times$. Triangles pointing down are 3~$\sigma$ upper limits and triangles pointing up are lower limits. For most of the lines, the $\pm$1~$\sigma$ error bars are smaller than the plot markers. }
    \label{fig:fluxes}
\end{figure*}

In Figure~\ref{fig:fluxes} we show the disk-integrated fluxes for molecules detected/not detected in the IRS~48 disk compared to the HD~100546 disk, which was observed as part of the same ALMA program \citep{Booth2023_hd100546}. For some molecules we detected multiple transitions but we only report the flux of a representative transition. These representative transitions are based on the strongest lines detected in the HD~100546 disk. 
In the case of CN, \ce{C_2H} and \ce{NO} the chosen lines are the strongest of the $N=3-2$, $N=4-3$ and $J=7/2-5/2$ hyper-fine groups, respectively. 
For \ce{SO_2} the $J=6_{(4,2)}-6_{(3,3)}$ is the strongest line detected and for SO the $J=7_8-6_7$ transition is the strongest. For \ce{CH_3OH} we pick the $J=7_{0}-6_{0}$ transition and for \ce{CH_3OCHO} and \ce{CH_3OCH_3} we use the $J=3_1-3_0$ and $J=19-18$ transitions which are both blends of multiple transitions. 
These fluxes are extracted from Keplerian masks which are 2\farcs0 and 4\farcs0 in radius for the IRS~48 and HD~100546 disks, respectively. If a molecule is undetected we give the 3~$\sigma$ upper limit on the flux where $\sigma$ is propagated from the rms in the channel maps and the number of pixels included in the mask \citep[e.g.,][]{2019A&A...623A.124C}. All of the line fluxes are listed in Table~\ref{tab:irs48_images} with their associated errors. After accounting for the different distances to the two sources (110~pc v 135~pc) HD~100546 is brighter in all of the lines aside from: \ce{SO}, \ce{^{34}SO}, \ce{SO_2}, \ce{NO}, \ce{H_2^{13}CO}, \ce{CH_3OH} and \ce{CH_3OCHO}. Note that the IRS~48 \ce{^{12}CO} (J=3-2) flux is a lower-limit due to foreground cloud absorption \citep[e.g., see Figure 2 in ][]{2014A&A...562A..26B}. 

To take into account the significantly different gas masses of the HD~100546 and IRS~48 disks (with HD~100546 $>$100$\times$ more massive than IRS~48; see Table~\ref{tab:A1}) we normalise the line fluxes with respect to the \ce{C^{17}O} $J=3-2$ line. The \ce{C^{17}O} line is the most optically thin CO isotopologue detected in both disks and this flux should be a good proxy for the total gas content in each disk \citep[e.g.,][]{2021ApJS..257....5Z}. These flux ratios are shown in Figure~\ref{fig:fluxes} and from this, there are significant differences in the relative intensities of the different molecular lines between these two disks. There are caveats to this comparison, e.g., if lines are optically thick in one or both of the disks and/or the excitation temperatures are very different. This is, however, a good starting point for comparing the two sources. The observed line strengths of most of the simple molecules are within a factor of 3 for the two disks. The differences in the line ratios become more significant when looking at the molecules that are already brighter in IRS~48. There is a factor 15 difference for NO and \ce{CH_3OCHO}, a factor 50 difference in the \ce{SO}, \ce{SO_2} and \ce{H_2^{13}CO} and a factor 80 difference for \ce{CH_3OH} between the two disks. The largest difference is in the \ce{^{34}SO}/\ce{C^{17}O} line ratio which is $\approx$130$\times$ higher in IRS~48 than HD~100546. 

\subsection{Column densities}

\begin{figure*}
    \centering
    \includegraphics[width=0.93\hsize]{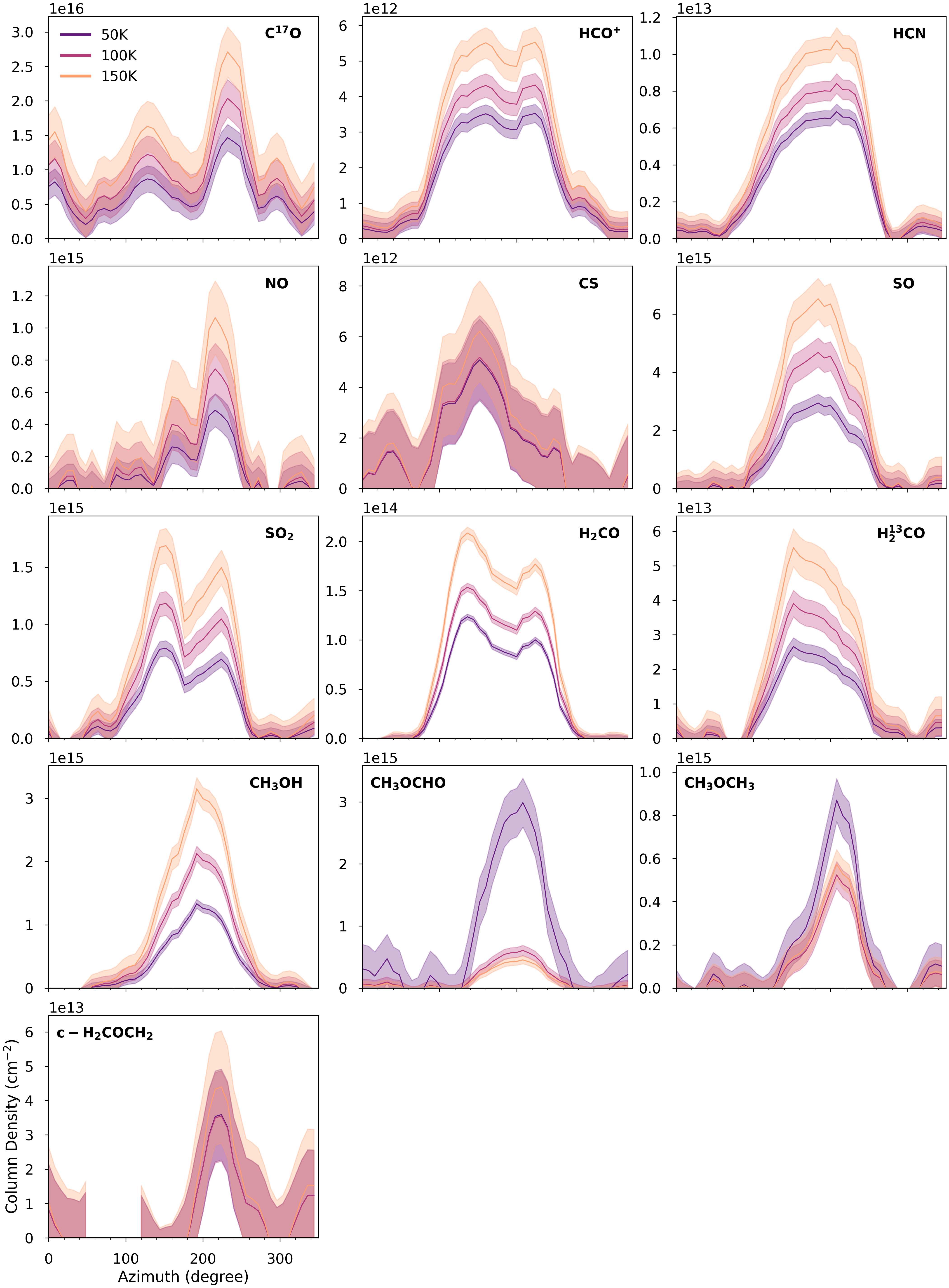}
    \caption{Azimuthal column density profiles for the IRS~48 disk determined at a range of assumed excitation temperatures.}
    \label{fig:irs48_azimuthal_columns}
\end{figure*}

We estimate column densities following the methods outlined in \citet{2018ApJ...859..131L} and in the same manner as \citet{Booth2023_hd100546}. For the molecules where multiple transitions are detected, e.g., \ce{CH_3OH} and \ce{SO_2}, we pick one representative transition. Future work will focus specifically on constraining the excitation conditions of these molecules individually. 
We compute azimuthal column density profiles for the IRS~48 disk from the profiles presented in Figure~\ref{fig:irs48_azimuthal_profiles} and explore a range of excitation temperatures: 50, 100 and 150~K. These temperatures are motivated by the observations and modelling results from \citet{2021A&A...651L...5V} and \citet{2023arXiv230300768L}. For the non-detected molecules, we calculate an upper limit propagated from the upper limits on the disk-integrated fluxes (listed in Table~\ref{tab:irs48_images}) assuming a conservative emitting area of a 2" aperture. The resulting profiles are shown in Figure~\ref{fig:irs48_azimuthal_columns} and the main results are as follows: 

\begin{itemize}
    \item The peak \ce{C^{17}O} column density is $\approx2.5\times10^{16}$ $\mathrm{cm^{-2}}$ at 100~K and the line is optically thin. 
    With the assumption of ISM isotope ratios, this is equivalent to a CO column density of $\approx5\times10^{19}$~$\mathrm{cm^{-2}}$.
    In Table~\ref{tab:irs48_peaks} we list the peak column density ratios of each of the molecules relative to the average CO column density across the IRS~48 disk. 
    
    \item The line emission from the simple molecules \ce{HCO^+}, \ce{HCN} and \ce{NO} are all optically thin. \ce{CN} and \ce{C_2H} are both undetected with the 3~$\sigma$ disk-averaged upper limit of CN $\leq2\times10^{12}$~$\mathrm{cm^{-2}}$ and \ce{C_2H} $\leq5\times10^{13}$~$\mathrm{cm^{-2}}$.
    
    \item The radical CS is detected with a peak column density of $\approx10^{13}~\mathrm{cm^{-2}}$ which is a factor of a few lower than the upper-limit report by \citet{2021A&A...651L...6B}. 
    \ce{H_2CS} is not detected with a column density upper limit of $<2\times10^{13}$~$\mathrm{cm^{-2}}$ which relative to CS is not constraining when comparing to other disks. 
    
    \item SO is abundant in the IRS~48 disk and therefore we use the $J=3_3-3_2$ transition for the column density calculation. This line has the lowest Einstein coefficient of the three SO transitions detected. The other two SO lines have lower column densities due to their higher optical depths. 
    This results in a peak column density of $5\times10^{15}~\mathrm{cm^{-2}}$ at 100~K. This is $\approx4\times$ higher than the \ce{SO_2} peak column density and results in a N(CS)/N(SO)$\approx10^{-3}$.
    When comparing the derived SO column density with the \ce{^{34}SO} column density, the ratio is consistent with 22, the local ISM \ce{^{32}S} to \ce{^{34}S} ratio \citep{1999RPPh...62..143W}, indicating the $J=3_3-3_2$ line is indeed optically thin. A detailed analysis on the S isotopes detected in these data will follow in a future work. OCS is not detected with a column density upper limit of $<10^{12}$~$\mathrm{cm^{-2}}$, less than a few percent of the SO column density. 
    
    \item Both \ce{H_2CO} and \ce{H_2^{13}CO} are robustly detected and we find a \ce{H_2CO}/\ce{H_2^{13}CO} column density ratio of $\approx4$. This is significantly lower than the expected \ce{^{12}C}/\ce{^{13}C} of 69 \citep{1999RPPh...62..143W} indicating optically thick \ce{H_2CO} emission or a lower isotope ratio.

    \item 
    The \ce{CH_3OH} column density at peaks at $\approx2\times10^{15} \mathrm{cm^{-2}}$.
    Using the column density derived for the main \ce{H_2CO} isotopologue results in a column density ratio of \ce{CH_3OH}/\ce{H_2CO} of 14$\pm1$ and using the \ce{H_2^{13}CO} and a C isotope ratio of 69 results in a ratio of 0.8$\pm0.1$. This means that if the \ce{H_2CO} is indeed optically thick the ratio of \ce{CH_3OH} to \ce{H_2CO} is $\approx$1. The \ce{CH_3OH} emission is still compact in these new data therefore, as discussed in \citet{2022A&A...659A..29B} the emission may be optically thick and beam diluted. This will be investigated further in Temmink et al. (in prep.) along with the constraints from \ce{{13}CH_3OH} which remains undetected.
    
    \item The peak abundance ratios of the COMs \ce{CH_3OCHO}, \ce{CH_3OCH_3} and \ce{c-H_2COCH_2} with respect to the peak \ce{CH_3OH} column density are $0.28\pm0.04$, $0.25\pm0.03$ and $0.017\pm0.006$, respectively, at a temperature of 100~K. \ce{CH_3CHO} is undetected with an upper-limit of $\approx4\times10^{13} \mathrm{cm^{-2}}$.
\end{itemize}

\begin{table*}
\caption{Ratios of the peak column density of different molecules (X) relative to disk averaged CO in the IRS~48 disk.}
    \centering
    \begin{tabular}{c  c c c } \hline \hline
      Molecule &  N(X)/N(CO) &  N(X)/N(CO) &  N(X)/N(CO) \\
      &  $T_{ex}$=50K& $T_{ex}$=100K &$T_{ex}$=150K  \\ 
    \hline
       \ce{HCO^+}   & $2.8\pm0.9\times10^{-7}$  &  $2.4\pm0.8\times10^{-7}$ &  $2.3\pm0.8\times10^{-7}$ \\
       \ce{HCN}     & $5.0\pm2.0\times10^{-7}$  & $5.0\pm2.0\times10^{-7}$  & $5.0\pm2.0\times10^{-7}$ \\
       \ce{CN}      & $<2.0\times10^{-8}$  & $<2.0\times10^{-8}$ & $<2.0\times10^{-8}$\\
       \ce{NO}      & $4.0\pm1.0\times10^{-5}$  & $4.0\pm2.0\times10^{-5}$ & $4.0\pm2.0\times10^{-5}$\\
       \ce{C_2H}    & $<5.0\times10^{-7}$  & $<5.0\times10^{-7}$ & $<5.0\times10^{-7}$\\
       \ce{CS}      & $4.0\pm2.0\times10^{-7}$ &  $3.0\pm1.0\times10^{-7}$ &  $3.0\pm1.0\times10^{-7}$ \\
       \ce{SO}      & $2.3\pm0.7\times10^{-4}$ & $2.6\pm0.8\times10^{-4}$& $2.7\pm0.9\times10^{-4}$ \\
       \ce{SO_2}    & $6.0\pm2.0\times10^{-5}$  & $6.0\pm2.0\times10^{-5}$ & $2.0\pm0.6\times10^{-5}$\\
       \ce{H_2CO}       & $1.0\pm0.3\times10^{-5}$  & $9.0\pm3.0\times10^{-6}$& $9.0\pm3.0\times10^{-6}$\\
        \ce{H_2^{13}CO} & $2.1\pm0.7\times10^{-6}$ & $2.1\pm0.7\times10^{-6}$ & $9.0\pm3.0\times10^{-7}$ \\
       \ce{CH_3OH}      & $1.1\pm0.3\times10^{-5}$  & $1.2\pm0.4\times10^{-5}$& $1.3\pm0.4\times10^{-5}$\\
        \ce{CH_3OCHO}   & $2.4\pm0.8\times10^{-4}$  & $3.0\pm1.0\times10^{-5}$   & $1.9\pm0.6\times10^{-5}$  \\
       \ce{CH_3OCH_3}   & $7.0\pm2.0\times10^{-5}$  & $2.9\pm0.9\times10^{-5}$  & $2.4\pm0.8\times10^{-5}$ \\
       \ce{c-H_2COCH_2} & $3.0\pm1.0\times10^{-6}$  & $2.0\pm1.0\times10^{-6}$ & $2.0\pm1.0\times10^{-6}$ \\
       \hline 
       \ce{CO} average & $1.3\pm0.4\times10^{19}$  & $1.8\pm0.4\times10^{19}$  & $2.3\pm0.8\times10^{19}$ \\
       \hline 
    \end{tabular}
     \tablecomments{
     These peak values are all taken at different radial and azimuthal locations and are all relative to the average CO column density derived from the \ce{C^{17}O} not the peak \ce{C^{17}O} column density.}
    \label{tab:irs48_peaks}
\end{table*}

\section{Discussion} \label{sec:discussion}

In this section, we discuss the physical and chemical origins of the observed molecular emission in the IRS~48 disk. We place this unique source in context with another chemically well-characterised protoplanetary disk namely, the HD~100546 disk which has been observed in the same frequecny setting within the same ALMA program. 

\subsection{The origin of the molecular sub-structures in the IRS~48 disk}

The simplest explanation for the molecular complexity and high relative column densities of oxygen-bearing volatiles in the IRS~48 disk is the sublimation of ices. 
With this new data, there are clear spatial offsets between the different molecules that complicate this picture. As seen in Figures~\ref{fig:mom02} and \ref{fig:irs48_azimuth},
the COMs have the most compact emission that peaks with the dust and these are also the species with the highest binding energies. The \ce{H_2CO} and HCN emissions are roughly co-spatial with a depression in both the \ce{H_2CO} and \ce{HCO^+} emissions where the COMs (and dust) emission is brightest.
Interestingly, the SO and \ce{SO_2} emissions, which \citet{2021A&A...651L...6B} proposed to originate first from the sublimation and photodissociation of \ce{H_2O} and \ce{H_2S} to OH and S respectively, are peaking radially further out in the disk compared to the \ce{CH_3OH}, the latter of which should trace the same region as the \ce{H_2O}. This may point to a different chemical origin for the SO and \ce{SO_2}. 
There may also be a link between gas leading and trailing in the Keplerian orbit of the dust trap. The orbital direction is highlighted in Figure~\ref{fig:irs48_azimuth}. 
The dust trap in the IRS~48 disk has been proposed to be a  large anticyclonic vortex \citep{2013Sci...340.1199V} therefore, it could be expected that there is additional radial and vertical mixing, or turbulence, and this will affect the disk chemistry. \citet{2011ApJS..196...25S} find that in their turbulent disk chemistry models the abundances of SO and \ce{SO_2} can increase by two orders of magnitude relative to the laminar disk due to the enhanced sublimation of ices. 
The interplay between the dust and line optical depth may also be influencing the observed emission structures. Therefore, a more detailed analysis of the IRS~48 line emission, including mapping the disk temperature structure, will be the focus of future work (Temmink et al. in prep).

\subsection{In context with other Herbig disks}

Both the IRS~48 and HD~100546 disks show rich reservoirs of complex organics and volatile sulphur that are yet to be detected in most other planet-forming disks (aside from HD~169142; \citealt{Booth2023}). 
The simplest explanation for the chemical origin of these species is via the sublimation of \ce{H_2O} and COM-rich ices. The brightness temperatures of the \ce{^{12}CO} in these disks are 100~K indicating the physical conditions for ice-sublimation are indeed possible \citep[see][]{2023A&A...670A.154W}. There are significant differences between the two disks, especially when considering their mass and size. The HD~100546 disk has a gas mass 500$\times$ and dust mass 100$\times$ higher than the IRS~48 disk (see Table~\ref{tab:A1}) and the HD~100546 CO disk extends to $\approx$600~au compared to $\approx$200~au for IRS~48. Given the different mass reservoirs in the disks, one may expect IRS~48 to have uniformly lower line fluxes than HD~100546 but, as shown in Figure~\ref{fig:fluxes}, this is not the case. On a disk average level the relative fluxes of simple oxygen molecules (NO, SO, \ce{SO_2}) and larger organics (\ce{H_2CO}, COMs) are 15 to 130 times brighter in IRS~48 than HD~100546. 

The \ce{HCO^+} abundance in the IRS~48 disk is similarly low as found in HD~100546 and HD~142527 which, is 2 orders of magnitude lower than found in the HD~163296 and MWC~480 disks \citep[see Table~4;][priv. comm. Temmink]{2021ApJS..257...13A, 2023A&A...675A.131T, Booth2023_hd100546}. This may be due to the low stellar X-ray flux of IRS~48 and/or the presence of gas-phase \ce{H_2O} (not yet detected in IRS~48 but only inferred, \citealt{2023arXiv230300768L}) which effectively destroys \ce{HCO^+}. We do not detect CN in IRS~48 and similar to HD~100546 it has a low CN/HCN ratio when compared to other disks. The low CS/SO ratio and non-detection of \ce{C_2H} is consistent with a disk C/O$<$1 as reported by \citet{2021A&A...651L...6B}. Similar to HD~100546, NO is the most abundant observed nitrogen carrier in the IRS~48 disk when compared to HCN or CN. The sulphur-bearing equivalent of \ce{H_2CO}, \ce{H_2CS}, was not detected in IRS~48. Given the high abundance of \ce{H_2CO} in IRS~48 this may be surprising but in the HD~100546 disk the \ce{H_2CS} follows the CS (as also found in MWC~480 and HD~169142; \citealt{2021ApJS..257...12L, Booth2023}) and not the sublimating SO in the inner disk. This indicates that the \ce{H_2CS} in disks is likely forming in the gas-phase at lower temperatures ($<$100~K) rather than having a significant abundance on the grains.

\subsection{Contextualising the volatile sulphur reservoir in the IRS~48 disk}

In IRS~48 SO, \ce{SO_2} and CS are detected but OCS and \ce{H_2CS} are not. With this family of molecules, we can compare the relative column density ratios of these species to both protostars and comets. In IRS~48 SO is the most abundant S-bearing volatile detected with peak column density ratios of \ce{SO_2}/SO of $\approx$26\%, CS/SO of $\approx$0.2\%, \ce{OCS}/SO of $<$0.3\% and, \ce{H_2CS}/SO of $<$1.0\%. The \ce{SO_2}/SO ratio in IRS~48 is similar to that detected in HD~100546 where again the SO column density is higher than the \ce{SO_2} column density \citep{Booth2023_hd100546}, but this is not the same as observed towards both protostars and comets. \citet{2018MNRAS.476.4949D} compare the volatile sulphur reservoirs in the comet 67P and towards the protostar IRAS~16293–2422~B. Comparing these environments to IRS~48: \ce{SO_2}, OCS and \ce{H_2CS} are all lower in abundance relative to SO in this disk than could be expected from the sublimation of cometary ices, although the SO/\ce{SO_2} ratio from 67~P has been shown to vary in time, exceeding 1 at points \citep{2016MNRAS.462S.253C}.
Additionally, \citet{2018MNRAS.476.4949D} show that OCS has strong variations between these two environments with OCS/SO $\approx$60\% in 67P and $\approx$560\% in IRAS~16293–2422~B, where in the latter source OCS is proposed to be enhanced due to UV irradiation. For both ratios, OCS would have been detectable in our data of the IRS~48 (and HD~100546) disk. \citet{2022ApJ...941...32B} find that the column density of OCS in the ices toward massive young stellar objects (MYSOs) correlates with the abundance of \ce{CH_3OH} ice. Therefore, with the detection of \ce{CH_3OH} in IRS~48 we may expect to also see OCS if these ices are dually inherited by the disk, but the binding energy of OCS (pure ice 2430~K, \citealt{2012MNRAS.425.1264W}) is significantly lower than than of \ce{CH_3OH} (on water ice 5000~K \citet{2020ApJ...904...11F, 2022ESC.....6..597M}). The median ice abundance of OCS relative to \ce{CH_3OH} towards the MYSOs target by \citet{2022ApJ...941...32B} is $\approx$1\% and in contrast for IRS~48 we find that gas-phase column density ratio of OCS/\ce{CH_3OH}$<$0.1\%. One explanation for the lack of OCS could be that during the disk lifetime, the volatile S in the simple inherited ices is converted to more refractory compounds like S allotropes due to processing via UV irradiation \citep{2022A&A...657A.100C}. 
Formation of S-allotropes can also can also act to destroy OCS on the ice, with models showing that OCS + S $\rightarrow$ \ce{S_2} + CO can be an important destruction pathway for OCS ice \citep{2019A&A...624A.108L}.
If \ce{S_2} is desorbed from grains it can also play an important role in gas-phase SO (and SO2) formation, via reactions with atomic O.
All in all, these comparisons show that the gas-phase volatile sulphur in IRS~48 is distinct to both the gas and ice detected towards protostars and in comets.





\subsection{Molecular complexity as evidence for ice processing?}

The degree of molecular complexity detected in the IRS~48 disk is unique for protoplanetary disks with three $\geq$7 atom COMs detected - \ce{CH_3OCHO}, \ce{CH_3OCH_3} and \ce{c-H_2COCH_2}. \ce{c-H_2COCH_2} is the first detection of a heterocyclic molecule in a protoplanetary disk. Heterocycles are abundant in comet 67P \citep{2023arXiv230800343H} and more generally these rings of carbon with an oxygen are of biological importance. The peak abundance ratios of these COMs with respect to the peak \ce{CH_3OH} column density show that these COMs have abundances $\approx$30, 25 and 2\% of \ce{CH_3OH}, respectively. Similarly, in the HD~100546 disk, in addition to \ce{CH_3OH}, \ce{CH_3OCHO} is also detected with an abundance of 70\% relative to \ce{CH_3OH}. Interestingly, \ce{CH_3OCH_3} is undetected in HD~100546 with an upper limit of $\lesssim$10\% relative to \ce{CH_3OH}. The slight differences in binding energies of \ce{CH_3OCHO} and \ce{CH_3OCH_3} are not sufficient to explain the lack of \ce{CH_3OCH_3} in HD~100546 since they are both lower than the binding energy of \ce{CH_3OH} \citep{2022ESC.....6..597M, 2023A&A...676A..80L}. 

Typically, in the warm gas around low and high-mass protostars these COMs have fractional abundances of a few percent of \ce{CH_3OH} \citep[e.g.,][]{2020A&A...635A..48M, 2020A&A...639A..87V, 2023arXiv230802688C}. The higher abundances we see in the Class II disks may simply be due to an underestimated \ce{CH_3OH} column density due to optically thick and beam diluted line emission. Deeper observations to target \ce{^{13}CH_3OH} isotopologues are needed to test this. Otherwise, if these high ratios are confirmed, these results reflect a different chemistry than is traced in observations of protostars. 
Similarly, the abundance ratio of \ce{CH_3OCHO} to \ce{CH_3OCH_3} has been shown to be remarkably constant across different evolutionary stages of star formation \citep{2020A&A...641A..54C,2023arXiv230802688C}. 
This ratio of $\approx$1 is also seen in IRS~48 but not for HD~100546 where we find a ratio $\gtrsim$7. \ce{c-H_2COCH_2} is an isomer of acetaldehyde (\ce{CH_3CHO}) and vinyl alcohol (\ce{CH_2CHOH}) both of which are undetected in our data. \ce{CH_3CHO} is typically the most abundant of these isomers by at least an order of magnitude: for example, observations of IRAS 16293-2422 find that \ce{CH_3CHO} is $\approx$10$\times$ more abundant than \ce{c-H_2COCH_2} and \ce{c-H_2COCH_2} relative to \ce{CH_3OH} is $\approx$0.05\% \citep{2017A&A...597A..53L, 2020A&A...635A..48M} whereas, in IRS~48, \ce{CH_3CHO}/\ce{c-H_2COCH_2} $\lesssim$1. 

The high abundance ratios of COMs with respect to \ce{CH_3OH} that we have observed so far in Class II disks and the variation in \ce{CH_3OCHO} and \ce{CH_3OCH_3} ratios between sources could be the result of the energetic processing of ices in disks. Over the millions of years that ices are present in disks they will be exposed to UV photons, X-ray and cosmic rays - especially if vertical mixing is prominent. These energetic processes can break apart \ce{CH_3OH} ice resulting in radicals (\ce{CH_3O}, \ce{HCO}, \ce{CH_3}) that can combine to form the more complex species \ce{CH_3OCHO}, \ce{CH_3OCH_3} and \ce{CH_3CHO} \citep{2009A&A...504..891O}. The specific branching ratios for these radicals will play a key role in setting the new COMs ice abundances \citep{2011ApJ...728...71L, 2014FaDi..168..389W}. \ce{c-H_2COCH_2} has been shown to form in the solid state via the reaction of \ce{C_2H_4} and O where \citet{2019ApJ...874..115B} find a branching ratio of 0.5 for \ce{c-H_2COCH_2} relative to \ce{CH_3CHO}. Given the upper limit on \ce{CH_3CHO} in the IRS~48 disk this may indicate that the formation of COMs via oxygen insertion reactions is also important. Finally, there may also be a non-negligible contribution from gas-phase reactions in the inner disk where the gas is warm $>$100~K, UV irradiated and at a significantly higher density than in protostellar envelopes. This needs to be tested with astrochemical models which we leave to further work. Additionally, a larger sample of disks is needed to understand the spread of COMs abundances in disks and better place IRS~48 in context. 
Upper limits on other COMs lines covered in these data and deuterated isotopes, e.g. \ce{HDCO}, in the IRS~48 disk, will be investigated in Kipfer et al. (in prep) where a further, more complete, comparison to proto-stellar environments and comets will be made. 


\section{Conclusion} \label{sec:conclusion}

This paper is the second in a series presenting an ALMA molecular line survey towards the disks around the Herbig Ae stars HD~100546 and IRS~48. Here we focus on the IRS~48 disk where we detect 16 different molecular species and our main results are as follows: 

\begin{itemize}
    \item We report the first robust detections of \ce{H_2^{13}CO}, \ce{^{34}SO}, \ce{^{33}SO} and \ce{c-H_2COCH_3} in protoplanetary disks and confirm the reported tentative detection of \ce{CH_3OCHO} from \citet{2022A&A...659A..29B} and \ce{CH_3OCH_3} is clearly seen. We also detect the simple molecules \ce{HCO^+}, \ce{HCN} and CS in the IRS~48 disk for the first time. 

    \item The IRS~48 disk hosts an extremely asymmetric dust trap in the south of the disk.  We find that all the molecular lines detected aside from CO show emission in the same region of the disk as the dust trap, including the simple molecules \ce{HCO^+}, \ce{HCN} and CS.

    \item The asymmetric molecular emissions from the different molecules are not all co-spatial. There are radial and azimuthal offsets in the peak position most clearly seen between the COMs and the SO and \ce{SO_2}. This warrants further investigation of the chemistry in turbulent vortices. 
        
    \item The low relative abundance of \ce{HCO^+} in IRS~48 is similar to the other Herbig disks HD~100546 and HD~142527, which could reflect the star's lower X-ray luminosity when compared to other sources.  Similar to regions of the HD~100546 disk, the CN/HCN ratio in IRS~48 is low $<$1, where the lack of CN may also be due to the low C/O ratio in the IRS~48 disk gas \citep{2023arXiv230300768L}. This is distinct from the elemental make up in the other Herbig Ae disks HD~163296 and MWC~480. 

    \item CS and HCN are the only molecules detected in the IRS~48 disk without oxygen and the low CS/SO ratio and the non-detection of \ce{C_2H} support the bulk of the gas in south of the IRS~48 disk having a C/O$<$1. In these data there is no evidence of an enhanced C/O$>$1 in the non-dust trap region of the disk. Further more, the partition of volatile S between SO, \ce{SO_2} and CS and, the non-detected OCS and \ce{H_2CS} is distinct to that measured for comets and protostars with OCS/SO $<$0.3\%.

    \item IRS~48 hosts the most chemically complex disk to date and the high abundances of COMs relative to \ce{CH_3OH} when compared to protostars as well as the different relative COMs ratios may indicate processing of the inherited ices in protoplanetary disks. The apparently high column density ratios of COMs to \ce{CH_3OH} needs to be confirmed via observations of optically thin tracers of \ce{CH_3OH}. i.e., the \ce{^{13}C} isotopologues.
\end{itemize} 

Our results solidify the IRS~48 disk as a unique astrochemical laboratory to study the full volatile reservoir available during planet formation and show the benefits of large unbiased surveys of protoplanetary disks. 
The clear association of the molecular emissions with the dust trap shows a strong coupling between the dust and ice chemistry. 
Nine different molecules have been detected for the first time in the IRS~48 disk in only two ALMA observing programs (2017.1.00834.S, 2021.1.00738) with just $\approx$10~hours of on-source time. The efficiency of these types of observations will improve dramatically with the planned Wideband Sensitivity Upgrade for ALMA which will increase both the simultaneously observable bandwidth and the imaging speed \citep{2023pcsf.conf..304C}. 

\begin{acknowledgements}
This paper makes use of the following ALMA data: 2021.1.00738S. We acknowledge assistance from Allegro, the European ALMA Regional Centre node in the Netherlands. ALMA is a partnership of ESO (representing its member states), NSF (USA) and NINS (Japan), together with NRC (Canada), MOST and ASIAA (Taiwan), and KASI (Republic of Korea), in cooperation with the Republic of Chile. The Joint ALMA Observatory is operated by ESO, AUI/NRAO and NAOJ. 
This work has used the following additional software packages that have not been referred to in the main text: Astropy, IPython, Jupyter, Matplotlib and NumPy \citep{Astropy,IPython,Jupyter,Matplotlib,NumPy}.
Astrochemistry in Leiden is supported by funding from the European Research Council (ERC) under the European Union’s Horizon 2020 research and innovation programme (grant agreement No. 101019751 MOLDISK).
A.S.B. is supported by a Clay Postdoctoral Fellowship from the Smithsonian Astrophysical Observatory. 
M.L. acknowledges support from grant 618.000.001 from the Dutch Research Council (NWO).
J.I.D. acknowledges support from an STFC Ernest Rutherford Fellowship (ST/W004119/1) and a University Academic Fellowship from the University of Leeds.
M.T. acknowledges support from the ERC grant 101019751 MOLDISK. 
C.W.~acknowledges financial support from the University of Leeds, the Science and Technology Facilities Council, and UK Research and Innovation (grant numbers ST/X001016/1 and MR/T040726/1).
L.E. acknowledges financial support from the Science and Technology Facilities Council (grant number ST/T000287/1).
S.N.~is grateful for support from RIKEN Special Postdoctoral Researcher Program (Fellowships), Grants-in-Aid for JSPS (Japan Society for the Promotion of Science) Fellows Grant Number JP23KJ0329, and MEXT/JSPS Grants-in-Aid for Scientific Research (KAKENHI) Grant Numbers JP 18H05441, JP20K22376, JP20H05845, JP20H05847, JP23K13155, and JP23H05441. 
Support for C.J.L. was provided by NASA through the NASA Hubble Fellowship grant No. HST-HF2-51535.001-A awarded by the Space Telescope Science Institute, which is operated by the Association of Universities for Research in Astronomy, Inc., for NASA, under contract NAS5-26555. 
\end{acknowledgements}

\bibliography{sample631}{}
\bibliographystyle{aasjournal}

\newpage
\appendix

\section{Observational set-up of IRS~48}

           
           


\begin{table}[h!]
    \centering
    \begin{tabular}{c c c c c c c c c c} \hline  \hline 
        Setting & Date & No. Antenna$^{*}$ & Integration Time& Baselines & Mean PVW & MRS &Phase & Flux/Bandpass\\
                 &     &   & (mins)  &  (m) & (mm)& (") & Calibrator & Calibrator \\ \hline
       A  & 30-05-2022 & 43 & 69 & 15.1-783.5 & 1.0 & 4.0 &  J1626-2951 & J1517-2422 \\
          & 08-06-2022 & 41 & 72 & 15.1-783.5 & 0.6 &  3.9 & J1626-2951 & J1427-4206\\ 
          & 08-06-2022 & 39 & 70 & 15.1-783.5 & 0.5 &  3.6 & J1626-2951 & J1427-4206\\ 
       B  & 28-05-2022 & 45 & 73& 15.1-783.5 & 1.2 &3.6 &  J1626-2951 & J1517-2422 \\
         & 28-05-2022 & 44 & 57 & 15.1-783.5 & 0.9 &3.6 &  J1626-2951 & J1517-2422 \\       
         & 28-05-2022 & 43 & 73& 15.1-783.5 & 0.9 & 4.0 & J1626-2951 & J1517-2422 \\  
         & 29-05-2022 & 44 & 73 & 15.1-783.5 & 1.4 & 3.9 & J1626-2951 & J1517-2422 \\   
         & 30-05-2022 & 43 & 73 & 15.1-783.5 & 0.9 & 4.1 &J1626-2951 & J1517-2422 \\     \hline \\
    \end{tabular}
    \caption{Execution block details. 
    $^{*}$ Number of antenna after flagging.}
    \label{tab:A3}
\end{table}

\newpage
\section{Molecular Data}

\begin{longtable*}{c c c c c c c}

    \caption{Molecular data of the transitions presented in this paper. This covers all of the molecules detected in the disk and particular non-detections of interest but not all of the transitions covered/detected. All data are taken from CDMS except for \ce{C^{17}O}, \ce{C_2H}, \ce{CH_3OCHO} and \ce{CH_3OCH_3} which are from JPL:      \citep{2016JMoSp.327...95E,1998JQSRT..60..883P}.} 
    \label{tab:lines}
    \\ \hline
    Molecule & Transition & Frequency (GHz) & E$_{\mathrm{up}}$ (K) & log10($\mathrm{A_{ul}}$) & $\mathrm{g_{u}}$ & Detection \\  \hline \hline
    \ce{^{12}CO}     & $J=3-2$ & 345.7959899  & 33.2 & -5.6027  & 7   & \cmark \\
    
    \ce{C^{17}O}     & $J=3-2$ & 337.0611298  & 32.7 & -5.6344  & 7   & \cmark \\
    
    \ce{HCO^+}       & $J=4-3$ & 356.7342230  & 42.8 & -2.4471  & 9   & \cmark \\
    
    \ce{HCN}         & $J=4-3$ & 354.5054779  & 42.5 & -2.6860  & 27  & \cmark \\
    \ce{H^{13}CN}    & $J=4-3$ & 345.3397693  & 41.4 & -2.7216  & 27  & - \\
    \ce{HC^{15}N}    & $J=4-3$ & 344.2001089  & 41.3 & -2.7258  & 9   &-\\
         CN      & $J=7/2-5/2, F=7/2-5/2$ & 340.2477700 & 32.7 &	-3.3839  & 10 & - \\
                     & $J=7/2-5/2, F=7/2-5/2$ & 340.2477700 & 32.7 & -3.4206  & 8  & - \\
                     & $J=7/2-5/2, F=5/2-3/2$ & 340.2485440 & 32.7 & -3.4347  & 6   & - \\
                    

    \ce{NO} & $J=7/2-5/2$,$\Omega=1/2- F=9/2-7/2$     & 351.0435240   &  36.1 & -5.2649  &  10    & \cmark \\
                     & $J=7/2-5/2$,$\Omega=1/2- F=7/2-5/2$     & 351.0517050   &  36.1 & -5.2662 &  8   & \cmark \\
                     & $J=7/2-5/2$, $\Omega=1/2- F=7/2-5/2$     & 351.0517050   &  36.1 & -5.3161 &  6   & \cmark \\

    \ce{HC_3N}        & $J=38-37$  & 345.6090100 & 323.5 &  -2.4812   & 77 & -\\
                      &  $J=39-38$	 & 354.6974631 & 340.5 &  -2.4473 & 79  & - \\

    \ce{CH_3CN}     &   $J=19_0-18_0 $& 349.4536999 & 167.7  & -2.5909  & 78 & - \\

                \ce{C_2H}       & $J=9/2-7/2,F=5-4$  & 349.3374558  & 41.9 & -3.7247 & 11   & \cmark \\
                     & $J=9/2-7/2,F=4-3$  & 349.3387284  & 41.9 & -3.7349 & 9   & \cmark \\


    \ce{CS}          & $J=7-6$  & 342.8828503 & 65.8 & -3.0774  & 15  & \cmark \\

    \ce{H_2CS}       & $J=10_{(1,10)}-9_{(1,9)}$ & 338.0831953  & 102.4     & 	-3.1995  & 63   & - \\ 
                     
    \ce{SO}          & $J=3_3-3_2$  & 339.3414590 & 25.5 & -4.8372   & 7  &  \cmark \\
                     & $J=7_8-6_7$  & 340.7141550 & 81.2 & -3.3023   & 15 &  \cmark \\ 
                     & $J=8_8-7_7$  & 344.3106120 & 87.5 & -3.2852   & 17 &  \cmark \\

    \ce{^{34}SO}     & $J=8_8-7_7$  & 337.5801467 & 77.3 & -3.3109 & 17 &  \cmark     \\
                     & $J=9_8-8_7$  & 339.8572694 & 86.1 & -3.2944 & 19 &  \cmark \\
                     
                    \ce{^{33}SO}     &$J=9_8-8_7$ $F=21/2-19/2$  & 343.0882949  & 78.0 &  -3.2819  & 22&  \cmark \\ 
                     &$J=9_8-8_7$ $F=19/2-17/2$  & 343.0880780  & 78.0 &  -3.2896   & 20&  \cmark \\ 
                     &$J=9_8-8_7$ $F=17/2-15/2$  & 343.0861019  & 78.0 & -3.2934    & 18 & \cmark \\ 
                     &$J=9_8-8_7$ $F=15/2-13/2$  & 343.0872979  & 78.0 &  -3.2916   & 16  & \cmark \\ 



                     &$J=7_8-6_7$ $F=17/2-15/2$  & 337.1986199  & 80.5 &  -3.3158  & 18&  \cmark \\ 
                     &$J=7_8-6_7$ $F=15/2-13/2$  & 337.1978453  & 80.5  &  -3.3283   & 16&  \cmark \\ 
                     &$J=7_8-6_7$ $F=13/2-11/2$  & 337.1980219  & 80.5 & -3.3351   & 14&  \cmark \\ 
                     &$J=7_8-6_7$ $F=11/2-9/2$  &  337.1993711  & 80.5  &  -3.3328   & 12 & \cmark \\ 

    \ce{SO_2}        &  $J=6_{(4,2)}-6_{(3,3)}$ &  357.9258478 & 58.6 & -3.5845  & 13 &  \cmark \\

    \ce{OCS}         & $J=28-27$  & 340.4492733 & 237.0 &  	-3.9378 & 57 &   - \\
                     & $J=29-28$  & 352.5995703 & 253.9 &  	-3.8918 & 59 &   -\\

    \ce{H_2CO}       & $J=5_{(1,5)}-4_{(1,4)}$ & 351.7686450  & 62.5 & 	-2.9201 & 33 &  \cmark\\
    
    \ce{H_2^{13}CO}  & $J=5_{(1,5)}-4_{(1,4)}$ & 343.3257130  & 61.3 &  -2.9517    & 33 &  \cmark \\
    \ce{CH_3OH}      & $J=7_{0}-6_{0}$ &  338.4086980& 65.0 &  -3.7691 & 60 & \cmark \\
    


        \ce{CH_3OCHO}    & $J=32_{(2,31)}-31_{(2,30)}$ & 344.0297653 &276.1 & -3.2099& 65 &  \cmark \\
 & $J=32_{(1,32)}-31_{(1,31)}$  &  344.0297645&276.1&-3.2099 & 65 &  \cmark \\
  &$J=32_{(0,32)}-31_{(0,31)}$  & 344.0295703 &276.1 & -3.2099& 65 &  \cmark \\
   & $J=32_{(1,32)}-31_{(1,31)}$  & 344.0295694 &276.1&-3.2099 & 65 &  \cmark \\


    \ce{CH_3OCH_3}   & $J=19_{(0,19)}-18_{(1,18)}$ AE & 342.6080601 &167.1& -3.2816 & 117 &  \cmark \\ 
 &  $J=19_{(0,19)}-18_{(1,18)}$ EA &  342.6080602& 167.1&  -3.2817& 78 &  \cmark\\ 
  & $J=19_{(0,19)}-18_{(1,18)}$ EE & 342.6081188 &167.1 & -3.2816 & 312 &  \cmark \\ 
   & $J=19_{(0,19)}-18_{(1,18)}$ AA & 342.6081774 &167.1 & -3.2816 & 195 &  \cmark \\ 



               \ce{CH_3CHO}         & $J=18_{(3,15)}-17_{(3,14)}$ A & 350.1334296 & 179.2& -2.82551&74  & - \\
                       & $J=18_{(3,15)}-17_{(3,14)}$ E & 350.1343816 & 179.2& -2.82596&74 & - \\

               
               \ce{c-H_2COCH_2}         & $J=11_{(1,10)}-10_{(2,9)}$ (ortho) &338.77197600   & 104.0 & -3.19217 & 69  & \cmark \\
                                        &$J=11_{(1,10)}-10_{(2,9)}$ (para) &338.77197600   & 104.0& -3.19212 & 115 & \cmark \\

    \hline
\end{longtable*}

\newpage
\section{Image properties}

\begin{table*}[h!]
    \caption{Properties of the line images for IRS~48 presented in Figure 2 and selected non-detections.}    \centering
    \begin{tabular}{c c c c c c c c c}
    \hline
    Molecule & Transition & robust & Beam & rms & Peak & Int. Flux \\  
    &  &  & ($\farcs \times \farcs$ ($^{\circ}$)) & (mJy beam $^{-1}$) &  (mJy beam $^{-1}$) & (mJy beam km s$^{-1}$) \\
    \hline \hline

    \ce{^{12}CO}    & $J=3-2$  & 0.5 & 0.34$\times$0.26~(-85.3)&0.94& 662.4& $>$25334.0\\
    \ce{C^{17}O}    & $J=3-2$  & 0.5 & 0.34$\times$0.28~(-89.6)&1.2&19.2&328.0$\pm$20.0\\
    \ce{HCO^+}      & $J=4-3$  & 0.5 & 0.33$\times$0.26~(-84.1)&1.19&30.8&529.0$\pm$21.0\\
    \ce{HCN}        & $J=4-3$  & 0.5 & 0.33$\times$0.26~(-84.8)&1.04&29.5&400.0$\pm$18.0\\
    \ce{H^{13}CN}   & $J=4-3$  & 0.5 & 0.34$\times$0.26~(-85.2)&0.92& -& $<$48.0\\
    \ce{HN^{15}N}   & $J=4-3$  & 0.5 & 0.34$\times$0.27~(-84.4)&0.99& -& $<$50.0\\
    \ce{CN}         & $N=4-3$  & 0.5 & 0.34$\times$0.27~(87.9)&1.03& - & $<$52.0\\
    
    \ce{HC_3N}      &$J=38-37$ & 0.5 &   0.34$\times$0.26~(-85.2)&0.94& - &$<66.0$\\
                    &$J=39-38$ & 0.5 &   0.33$\times$0.26~(-84.7)&1.04&  - &$<75.0$\\
                    
    \ce{CH_3CN}      &$J=19_0=18_0$ & 0.5 & 0.33$\times$0.26~(-89.3)&1.05& - &$<$75.0\\

    \ce{NO}         & $J=7/2-5/2$ & 0.5 & 0.33$\times$0.26~(-89.9)&1.08& 8.16 &127.0$\pm$25.0\\
    \ce{C_2H}       & $N=4-3$  & 0.5 & 0.33$\times$0.26~(-89.4)&1.02& - &$<$53.0\\
    
    \ce{CS}         & $J=7-6$ & 0.5 &  0.34$\times$0.27~(-84.4)&1.03& 5.82 & 40$\pm$17.0\\
    
    \ce{H_2CS}      &$J=10_{(1,10)}-9_{(1,9)}$& 0.5 &  0.34$\times$0.27~(-89.3)&1.16& - &$<$58.0\\
    \ce{SO}         &$J=3_3-3_2$ & 0.5 &    0.34$\times$0.27~(87.9)&0.79& 87.60&1070.0$\pm$13.0\\
                    &$J=7_8-6_7$ & 0.5 & 0.34$\times$0.27~(-85.3)&1.0& 84.96 &1063.0$\pm$23.0\\
                    &$J=8_8-7_7$ & 0.5 &   0.34$\times$0.27~(88.2)&1.05& 27.85 &232.0$\pm$24.0\\

    \ce{^{34}SO}    & $J=8_8-7_7$ & 0.5 &  	0.34$\times$0.27~(88.0)&0.98& 28.81&251.00$\pm$22.0\\
                    & $J=9_8-8_7$ & 0.5 &   	0.34$\times$0.27~(-89.4)&1.13&23.93&187.0$\pm$26.0\\

    \ce{^{33}SO}    &$J=7_8-6_7$ & 0.5 &   0.34$\times$0.27~(-84.5)&1.0& 10.64 &115.0$\pm$23.0\\
                    &$J=9_8-8_7$ & 0.5 &  0.34$\times$0.28~(-89.5)&1.15& 8.60&60.0$\pm$20.0\\

    \ce{SO_2}       &$J=6_{(4,2)}-6_{(3,3)}$ & 0.5 & 0.33$\times$0.26~(-84.2)&0.88&26.83 &261.0$\pm$15.0\\

    \ce{OCS}   & $J=28-27$ & 0.5 & 0.34$\times$0.27~(88.0)&0.98& - &$<$67.0\\
                 & $J=27-26$        & 0.5 &  0.33$\times$0.26~(-90.0)&1.42& - &$<$101\\

    \ce{H_2CO}      &$J=5_{(1,5)}-4_{(1,4)}$ & 0.5 & 0.33$\times$0.26~(90.0)&1.21&77.54&824.0$\pm$21.0\\
    
    \ce{H_2^{13}CO} & $J=5_{(1,5)}-4_{(1,4)}$&0.5 &  0.34$\times$0.27~(-84.5)&1.02&17.34&169.0$\pm$17.0\\
    
    
    
    \ce{CH_3OH}     & $J=7_{0}-6_{0}$ & 0.5& 0.34$\times$0.27~(-89.5)&1.11& 19.83 &182.0$\pm$18.0\\

    \ce{CH_3OCHO}   & $J=31-30$ & 0.5& 0.34$\times$0.27~(-84.4)&0.98& 12.50 & 95.0$\pm$23.0 \\
    
    \ce{CH_3OCH_3}& $J=19-28$ & 0.5   & 0.34$\times$0.27~(-89.6)&0.82& 13.14 & 87.0$\pm$23.0\\
    
    \ce{CH_3CHO}   & $J=18_{(3,15)}-17_{(3,14)}$& 0.5& 	0.34$\times$0.27~(-89.7)&0.79& -& $<$75\\ 
    
   \ce{c-H_2COCH_2}& $J=11_{(1,10)}-10_{(2,9)}$& 2.0 &  0.37$\times$0.3~(27.7)&0.77& 5.70& $39.0\pm$24  \\

    \hline
    \end{tabular}
    \label{tab:irs48_images}
\end{table*}

\newpage
\section{Full spectrum matched filter response}

\begin{figure}[h!]
    \centering
    \includegraphics[width=0.9\hsize]{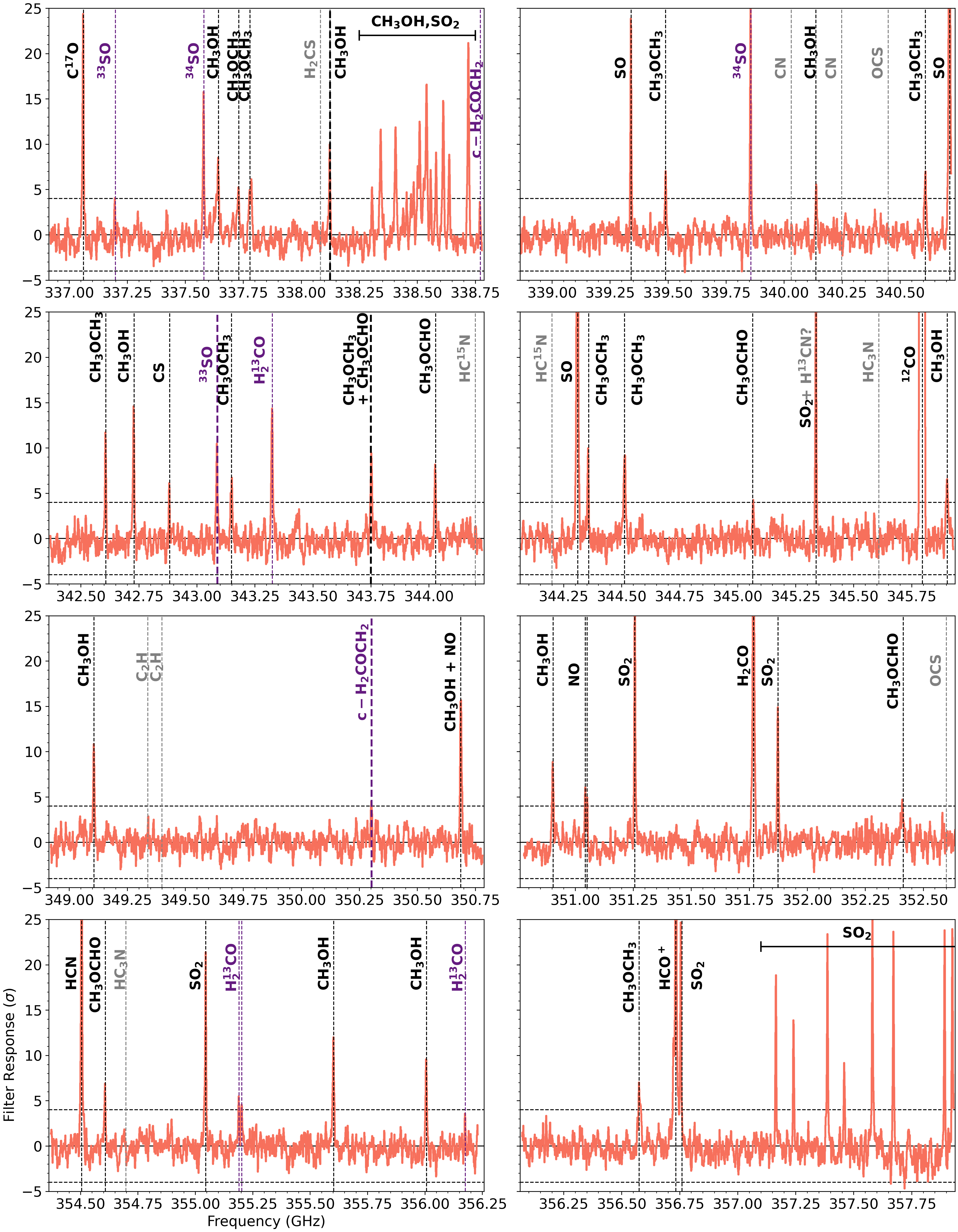}
    \caption{IRS~48 matched filter response using a 150~au in radius Keplerian model. Detected molecules/transitions above the 4$\sigma$ level are labelled. New disk molecules are noted in purple and notable non-detections are shown in grey.}
    \label{fig:irs48_filter}
\end{figure}

\end{document}